\documentclass[royal]{witpress}
\usepackage{graphicx}

\begin{document}

\title{2D BEM modeling of a singular thermal diffusion free boundary problem with phase change}

\author{V. S. Nikolayev \& D. A. Beysens}

\address{ESEME, Service des Basses Temp\'eratures, DSM/DRFMC, CEA-Grenoble, France\thanks{Mailing address:
CEA-ESEME, Institut de Chimie de la Mati\`ere Condens\'{e}e de Bordeaux,
  87, Avenue du Dr. Schweitzer, 33608 Pessac Cedex, France}}

\maketitle

\begin{abstract}
  We report a 2D BEM modeling of the thermal diffusion-controlled growth of a vapor bubble attached to a heating
surface during saturated pool boiling. The transient heat conduction problem is solved in a liquid that
surrounds a bubble with a free boundary and in a semi-infinite solid heater. The heat generated
homogeneously in the heater causes evaporation, i. e. the bubble growth. A singularity exists at the point
of the triple (liquid-vapor-solid) contact. At high system pressure the bubble is assumed to grow slowly,
its shape being defined by the surface tension and the vapor recoil force, a force coming from the liquid
evaporating into the bubble. It is shown that at some typical time the dry spot under the bubble begins to
grow rapidly under the action of the vapor recoil. Such a bubble can eventually spread into a vapor film
that can separate the liquid from the heater, thus triggering the boiling crisis (Critical Heat Flux
phenomenon).
\end{abstract}

\section{Introduction}

Boiling is widely used to transfer heat from a solid heater to a liquid. The bubble growth in boiling
attracted much of attention from many scientists and engineers. In spite of these efforts, some important
aspects of growth of a vapor bubble attached to a solid heater remain misunderstood even on a
phenomenological level. The most important aspect is the boiling crisis, a transition from nucleate
boiling (where vapor bubbles nucleate on the heater) to film boiling (where the heater is covered by a
continuous vapor film). The boiling crisis is observed when the heat flux $q_S$ from the solid heater
exceeds a threshold value which is called the "Critical Heat Flux" (CHF). The rapid formation of the vapor
film on the heater surface decreases steeply the heat transfer efficiency and leads to a local heater
overheating. In the industrial heat exchangers, the boiling crisis can lead to melting of the heater thus
provoking a dangerous accident. Therefore, the knowledge of the CHF is extremely important. However, the
CHF depends on many parameters. At this time, there are several semi-empirical correlations that predict
the CHF more or less reliably for several particular regimes of boiling and heater configurations, see
\cite{Dhir} for a recent review. However, a clear understanding of the triggering mechanism of the boiling
crisis is still lacking.

The knowledge about what happens at the foot of the bubble which grows attached to the heater is crucial
for the correct modeling of the boiling crisis. Unfortunately, the experimental observations at large heat
fluxes close to the CHF are complicated by the violence of boiling and optical distortions caused by the
strong temperature gradients. We proposed recently \cite{PRE} to carry out boiling experiments in the
proximity of the critical point where the CHF is very small and the bubble evolution is very slow.
However, microgravity conditions are necessary in this case to obtain a convex bubble shape in order to
observe a behavior similar to the terrestrial boiling.

The bubble foot contains the contact line of the bubble with the heater. This triple solid-liquid-gas
contact line is a line of singularity points both for the hydrodynamic (see \cite{JFM} and refs. therein)
and for the heat conduction problems. In the present article we consider only the heat conduction part by
assuming the slow growth and the quasi-static bubble shape which is common for the high pressure boiling.
The results of such a calculation have already been described in \cite{IJHMT}. The present article deals
with the problem framework and some calculation details related to BEM.

\section{Boundary conditions for the contact line problem}\label{sec2}

\begin{figure}
  \begin{center}
  \includegraphics[height=6cm]{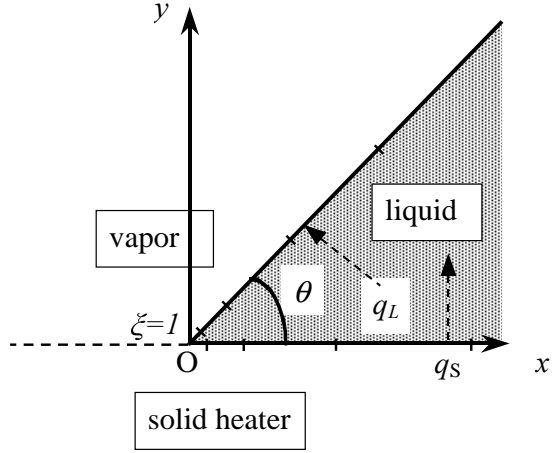}
  \end{center}
\caption{Geometry for the calculation of the heat conduction in the wedge geometry. The BEM discretization
of the wedge is also illustrated.\label{Fig1}}
\end{figure}

The choice of the boundary conditions adopted in the contact line is very important. Since the contact
line is triple, boundary conditions should be specified at three surfaces that intersect there
(Fig.~\ref{Fig1}).
For the gas-liquid interface, we adopt the constant-temperature boundary condition with the temperature
that corresponds to the saturation temperature $T_{sat}$ for the given (constant) system pressure. The gas
phase is assumed to be non-conductive, the heat flux through the solid-gas interface being zero. The
boundary condition on the solid-liquid interface remains to be defined. It is the subject of the rest of
this section.

There are three kinds of boundary conditions. Let us consider them on an example of the 2D wedge geometry
as illustrated in Fig.~\ref{Fig1} for which some important solutions can be obtained in analytical form.

Since the heater is a far better heat conductor than the liquid, the constant temperature boundary
condition ($T=T_S=$const along the solid-liquid interface) seems natural. In order to maintain boiling,
$T_S>T_{sat}$ should be satisfied. The resulting problem is ill posed because the temperature is
discontinuous along the boundary of the liquid domain at the contact point O in Fig.~\ref{Fig1}. This
discontinuity leads to a singular behavior of the heat flux $q_L$ through the gas-liquid boundary,
$q_L(y)\propto y^{-1}$ \cite{Carslaw} and is not integrable. Note that the integral $\int q_L {\rm d}y$ is
very important because it defines the amount of liquid evaporated into the bubble and thus the bubble
growth rate, see Eq.~(\ref{eqnV}) below. Since the result is infinite, the first kind boundary condition
cannot be used.

\begin{figure}[htb]
  \begin{center}
  \includegraphics[height=6cm]{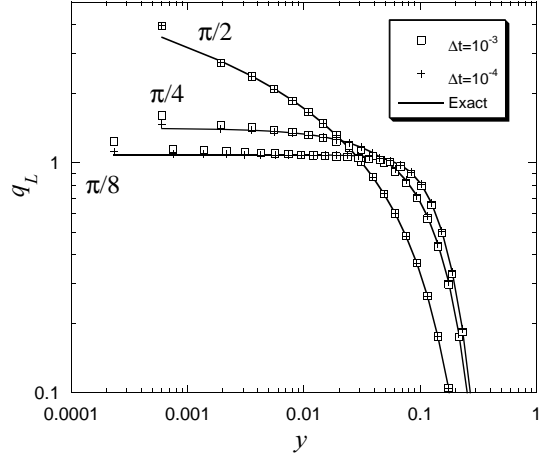}
  \end{center}
\caption{The $q_L(y)$ curves calculated for the $\pi/2$, $\pi/4$ and $\pi/8$ wedges for the values of the
parameters $q_0=1$, $\alpha_L=1$, $t=0.01$ and $d_{min}=0.001$ and two values of the time step $\Delta t$.
The results of the numerical solution by BEM (data points) should be compared with the exact analytical
solutions (lines).} \label{test}
\end{figure}

Another choice is the constant heat flux $q_S$ along the solid-liquid interface, which can be reasonable
for a thin heater. The resulting transient problem can be solved analytically in the liquid domain by the
reflections method \cite{Carslaw}. Its solution can be obtained in the closed form for several contact
angles $\theta$, see Fig.~\ref{Fig1}. The solutions obtained in \cite{IJHMT} for $\theta=\pi/4$ and
$\theta=\pi/8$ result in a constant value for $q_l(y\rightarrow 0)$. Although the solution for
$\theta=\pi/2$ diverges $q_L(y)\propto \log(-y)$ \cite{EuLet}, it is integrable. In spite of these
advantages, the constant heat flux boundary condition is not suitable for the bubble growth problem
because it cannot be used in the dry spot (i. e. solid-gas contact) area where the heat flux should be
zero. However, these analytical solutions can be used to test the BEM solver code (see Fig.~\ref{test}).

A remaining option is the boundary conditions of a third kind, i.e. the coupling of the temperatures and
the fluxes at the solid surface. The heat conduction problem is required to be solved in the solid domain
in addition to the liquid domain. Unfortunately, we cannot solve the problem analytically in this case.
Qualitatively, one can expect an integrable divergence of $q_L(y)$ which should appear because of the
influence of the solid-gas contact area adjacent to the contact line. Since the heat flux that comes from
the bulk of the solid heater is not able to pass through this area, it should be necessarily deviated
towards the neighboring solid-liquid contact area thus increasing $q_S$ at small $x$. Since $q_S\approx
q_L$ near the contact line, $q_L(y)$ should vary steeper near the contact line than in the constant $q_S$
case and is likely to diverge.

One can argue that the necessity of the calculation of the temperature field in the heater is a heavy
complication that justifies the approximation of the simultaneous application of the boundary conditions of
constant heat flux outside the dry spot and zero heat flux inside. However, the above considerations show
that the behavior of $q_L(y)$ can deviate from its real behavior even qualitatively. Since such a large
error cannot be admitted in the calculation of $q_{L}$ that strongly influences the bubble dynamics, we
need to calculate rigorously the conjugate heat conduction problem.

\section{Mathematical problem statement}

\begin{figure}[htb]
  \begin{center}
  \includegraphics[height=6cm]{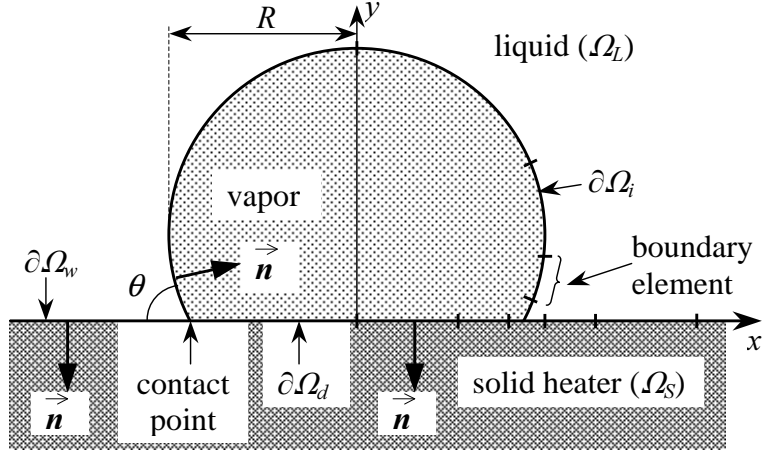}
  \end{center}
\caption{Vapor bubble on the heater (domain $\Omega_S$) surrounded by the liquid (domain $\Omega_L$). The
chosen direction of the unit normal vector $\vec{n}$ is shown for each of the subcontours
$\partial\Omega_w$, $\partial\Omega_d$ and $\partial\Omega_i$. The discretization is illustrated for the
right half of the subcontours. (From \cite{IJHMT} with permission from Elsevier Science) } \label{bubble}
\end{figure}

The 2D heat conduction problem in the domain $\Omega_L\cup\Omega_S$ (see Fig.~\ref{bubble})
reads
\begin{eqnarray}
{\partial T_L\over\partial t}=\alpha_L \nabla^2 T_L,\quad\vec{r}\in\Omega_L\label{eqnL}\\
{\partial T_S\over\partial t}=\alpha_S \nabla^2 T_S+{\alpha_S\over k_S}j(t), \quad\vec{r}\in\Omega_S
\label{eqnS}
\end{eqnarray}
where $\alpha$ and $k$ is the thermal diffusivity and conductivity respectively, the indices L and S
identify the liquid and the solid heater, and $\vec{r}=(x,y)$ denotes a point. The heat is assumed to be
generated homogeneously in the heater with the power $j(t)$ per unit volume. We choose $j(t)=C/\sqrt{t}$,
where $C$ is constant. This condition results in a constant in time value of $q_S$ far from the bubble,
see (\ref{q0}) below. Initially $(t=0)$ the temperature is homogeneous $T_L=T_S=T_{sat}$. The vapor bubble
is assumed to be already nucleated. Since we assume the zero contact angle $\theta=0$, the bubble shape is
spherical with the radius $R_0$. The boundary conditions are formulated on the moving gas-liquid interface
$\partial\Omega_i$ $(\left.T_L\right|_{\partial\Omega_i} =T_{sat})$, on the solid-liquid interface
$\partial\Omega_d$ $(\left.\partial T_L/\partial y\right|_{\partial\Omega_d}=0)$ and on the solid-liquid
interface $\partial\Omega_w$:
\begin{equation}
\begin{array}{l}
\displaystyle q_S=-k_S\left.{\partial T_S\over\partial y}\right|_{\partial\Omega_w}=
 -k_L\left.{\partial T_L\over\partial y}\right|_{y=0},\\
\left.T_S\right|_{\partial\Omega_w}=\left.T_L\right|_{y=0}.
\end{array}
\label{coupl}
\end{equation}
Because of the axial symmetry, this problem needs to be solved only for $x>0$.

The shape of the gas bubble is calculated from the quasi-static equation \cite{EuLet}
\begin{equation} K(\vec{r}_i)\sigma=\lambda+P_r(\vec{r}_i),\label{surf}
\end{equation}
where $K$ is the curvature of the bubble at the point on the surface $\vec{r}_i=(x_i,y_i)$, $\sigma$ is the
vapor-liquid interface tension and $\lambda$ is a constant difference of pressures between the vapor and
the liquid. The vapor recoil pressure
\begin{equation}\label{Pr}
  P_r(\vec{r}_i)=[q_L(\vec{r}_i)/H]^2(\rho_V^{-1}-\rho_L^{-1}),
\end{equation}
where $q_L=-k_L(\vec{n}\cdot\nabla) \left.T_L\right|_{\partial\Omega_i}$ appears due to the uncompensated
momentum of vapor molecules that leave the interface. The latent heat of vaporization is denoted by $H$,
$\rho_V$ and $\rho_L$ being the vapor and liquid densities. Eq. (\ref{surf}) is convenient to solve when
the bubble contour is described in parametric form $\vec{r}_i=\vec{r}_i(\xi)$ where $\xi$ is the distance
from the topmost point of the bubble to the point $\vec{r}_i$ measured along the bubble contour. When $\xi$
is non-dimensionalized by the half-length of the bubble contour $L$, $\xi=1$ corresponds to the contact
point  and (\ref{surf}) becomes equivalent to the following set of ODEs \cite{IJHMT}:
\begin{eqnarray}
{\rm d}x_i/{\rm d}\xi  &=&L\cos u,  \label{vol1} \\ {\rm d}y_i/{\rm d}\xi  &=&-L\sin u,  \label{vol2} \\
{\rm d}u/{\rm d}\xi &=&L(\lambda +P_{r}(\xi ))/\sigma   \label{vol3}
\end{eqnarray}
were $u=u(\xi)$ is an auxiliary function. The boundary conditions for this set read $x_i(0)=0$, $u(0)=0$,
$y_i(1)=0$. By fixing the contact angle $u(1)=\pi-\theta$ one can determine the unknown $L$ from
(\ref{vol3}),
\begin{equation} L=(\pi-\theta)\sigma\left[\int\limits_0^1 P_r(\xi) {\rm d}\xi
+\lambda\right]^{-1},\label{L}
\end{equation}
$\theta=0$ is assumed in the rest of this article. The constant $\lambda$ should be determined using the
known volume $V$ of the 2D bubble,
\begin{equation}
V=-{1\over 2}\int\limits_{(\partial\Omega_i)} (x_in_x+y_in_y) \;{\rm d}\,\partial\Omega,\label{V}
\end{equation}
where $n_x$ and $n_y$ are the components of $\vec{n}$, Fig.~\ref{bubble}. The bubble volume increases in
time due to evaporation at $\partial\Omega_i$
\begin{equation}
H\rho_V{{\rm d}V\over{\rm d}t}=\int\limits_{(\partial\Omega_i)} q_L(\vec{r}_i) \;{\rm d}\,\partial\Omega,
\label{eqnV}
\end{equation}
This equation is used widely to describe the thermally controlled bubble growth. The initial condition is
$V(t=0)=4/3\,\pi R_0^3$.

The problem (\ref{eqnL}-\ref{eqnV}) is now complete. It can be solved by BEM. However, it is not
convenient to solve by BEM because the temperature and its gradient are both non-zero at infinity (more
precisely, at $x\rightarrow\infty$), where the closing subcontours for the domains $\Omega_L$ and
$\Omega_S$ are located. We solve this problem easily by subtracting the solutions at $x\rightarrow\infty$.
These solutions for the both domains read \cite{IJHMT}
\begin{eqnarray}
T_L^{inf}&=& T_{sat}+{q_0\over k_L}\left[\sqrt{4\alpha_L t\over\pi}\exp\left(-{y^2\over 4\alpha_L
t}\right)-
y\,\mbox{erfc}\left({y\over 2\sqrt{\alpha_Lt}}\right)\right], \label{TLinf} \\
T_S^{inf}&=& T_{sat}+{2\alpha_S\over k_S}C\sqrt{t}-\nonumber\\&{q_0\over k_S}&\left[\sqrt{4\alpha_S
t\over\pi}\exp\left(-{y^2\over 4\alpha_S t}\right)+ y\,\mbox{erfc}\left(-{y\over
2\sqrt{\alpha_St}}\right)\right], \label{TSinf}
\end{eqnarray}
where erfc$(\cdot)$ is the complementary error function \cite{Abr}. As one can easily find, the flux from
the heater $q_0$ far from the bubble
\begin{equation}
q_0\equiv q_S(x\rightarrow\pm\infty)=C\sqrt{\pi}\alpha_S k_L/(k_S\sqrt{\alpha_L}+k_L\sqrt{\alpha_S}),
\label{q0}
\end{equation}
is constant in time. We use its value as a control parameter instead of $C$.

By introducing the characteristic scales for time ($\Delta t$, the time step), length ($R_0$, the initial
bubble radius), heat flux ($\bar{q}$), and thermal conductivity ($\bar{k}$), all other variables can be
made non-dimensional. In particular, the characteristic temperature scale in the system is
$\bar{q}R_0/\bar{k}$. The following four non-dimensional groups define completely the behavior of the
system
\begin{eqnarray*}
{\rm Fo}_{L,S}=\alpha_{L,S}\Delta t/R_0^2 \mbox{ --- Fourier numbers},\\ {\rm
Ja}={k_L\bar{q}R_0\over\rho_VH\bar{k}\alpha_L} \mbox{
--- Jakob number},\\ {\rm Hi}={R_0\,\bar{q}^2\over\sigma
H^2}(\rho_V^{-1}-\rho_L^{-1}) \mbox{
--- Hickman number},
\end{eqnarray*}
together with the non-dimensionalized values of $q_0$ and $k_{L,S}$. The non-dimensional heat transfer
problem can now be formulated in terms of temperatures $\psi_{L,S}=(T_{L,S}-T_{L,S}^{inf})/(\bar{q}R_0/
\bar{k})$ and fluxes $\zeta_{L,S}=k_{L,S}\,\partial\psi_{L,S}/\partial\vec{n}$.

\section{Numerical implementation}

\subsection{BEM formulation}

Since $\psi_{L,S}=\zeta_{L,S}=0$ at infinity, the heat conduction problem is equivalent to a set of two BEM
equations \cite{Wrobel} for the open integration contours
$\partial\Omega_L=\partial\Omega_i\cup\partial\Omega_w$ and
$\partial\Omega_S=\partial\Omega_d\cup\partial\Omega_w$.  Using $t=0$ as the initial time moment and
taking into account that $\psi_{L,S}(t=0)=0$ so that the volume integral disappears, these equations read
\begin{eqnarray}
\int\limits_{0}^{t_F}&{\rm d}t\int\limits_{(\partial\Omega_{L,S})}\Biggl[G^{L,S}(\vec{r'},t_F;\vec{r},t)
\left({\rm Fo}_{L,S}{\zeta_{L,S}(\vec{r},t)\over
k_{L,S}}+\psi_{L,S}(\vec{r},t) v^n(\vec{r},t) \right)- \nonumber\\
&\left.{\rm Fo}_{L,S}\, \psi_{L,S}(\vec{r},t){\partial_r
G^{L,S}(\vec{r'},t_F;\vec{r},t)\over\partial\vec{n}_r} \right]{\rm d}_r\partial\Omega={1\over
2}\psi_{L,S}(\vec{r'},t_F),\label{Ieq}
\end{eqnarray}
where the point $\vec{r'}$ belongs to $\partial\Omega_{L,S}$ respectively, $v^n$ is the normal component
of the contour velocity (nonzero only on $\partial\Omega_i$) and
\begin{equation}
G^{L,S}(\vec{r'},t_F;\vec{r},t)={1\over 4\pi{\rm Fo}_{L,S}(t_F-t)}\exp\left[-{|\vec{r'}-\vec{r}|^2\over
4\,{\rm Fo}_{L,S}(t_F-t)}\right].\end{equation} In the following, the indices $L$ and $S$ for all
variables in the equations will be dropped for the sake of clarity.

The constant element BEM \cite{Breb} was used, i. e. both $\zeta$ and $\psi$ were assumed to be constant
during any time step and on any element $\partial\Omega_j$, their values on the element being associated
with the values at the node $B_j$ in the middle of the element approximated by a straight segment that
joins its ends $M_{j-1}$ and $M_j$. The time steps are equal, i.e. $t_f=f$. The values of $\zeta$ and
$\psi$ on the element $j$ at time $f$ are denoted by $\zeta_{fj}$ and $\psi_{fj}$. Each of the integral
equations (\ref{Ieq}) reduces to the system of linear equations
\begin{equation}
\sum\limits_{f=1}^F\sum\limits_{j=1}^{2N_f}[(\zeta_{fj}/k+ \psi_{fj}v^n_{fj}/{\rm
Fo})G_{ij}^{Ff}-\psi_{fj}H_{ij}^{Ff}]=\psi_{Fi}/2,\label{lin1}
\end{equation}
where $N_f$ is the number of elements on one half of the integration contour at time step $f$, $F_{max}$ is
the maximum calculation time; $i=1\ldots 2N_F$ and $F=1\ldots F_{max}$; $H_{ij}$ and $G_{ij}$ are the BEM
coefficients:
\begin{equation}
\begin{array}{l}
\displaystyle G_{ij}^{Ff}={\rm Fo}\int\limits_{f-1}^{f}{\rm
d}t\int\limits_{(\partial\Omega_j)}G(\vec{r}_i,F;\vec{r},f){\rm d}_r\partial\Omega,\\\displaystyle
H_{ij}^{Ff}={\rm Fo}\int\limits_{f-1}^{f}{\rm d}t\int\limits_{(\partial\Omega_j)}{\partial
G(\vec{r}_i,F;\vec{r},f)\over\partial\vec{n}_r}{\rm d}_r\partial\Omega.
\end{array}
\label{ght}\end{equation} Since the calculation of these coefficients takes the most of computation time,
it should be made fast.

\subsection{Algorithm for the BEM coefficients}

 The value of each particular BEM coefficient for the element $\partial\Omega_j=(M_{j-1},M_j)$ of the length $l$ is
calculated using the coordinate transformation \cite{Hess} to the Cartesian system where $B_j$ is the
reference point and the $x$ axis is directed toward $M_{j-1}$. The direction of the normal vector $\vec{n}$
coincides with the $y$ axis. The time integration \cite{Breb} results in
\begin{eqnarray}
G_{ij}^{Ff}&=&\int\limits_{-l/2}^{l/2}{1\over 4\pi}\left[E_1\left({(x+u)^2+y^2\over 4\,{\rm
Fo}(F-f+1)}\right)-E_1\left({(x+u)^2+y^2\over 4\,{\rm
Fo}(F-f)}\right)\right]{\rm d}u,\label{gfull}\\
H_{ij}^{Ff}&=&\int\limits_{-l/2}^{l/2}{y\over
2\pi}\left[{\exp\left(-{(x+u)^2+y^2\over 4\,{\rm
Fo}(F-f+1)}\right)\over(x+u)^2+y^2}-{\exp\left(-{(x+u)^2+y^2\over
4\,{\rm Fo}(F-f)}\right)\over(x+u)^2+y^2}\right]{\rm
d}u,\label{hfull}
\end{eqnarray}
where $(x,y)$ denote the coordinates of the point $\vec{r}_i$ in the new reference system and $E_1(\cdot)$
is the exponential integral \cite{Abr}. The situation $i=j$ (i.e. where $x=y=0$) is particular, which is a
quite general feature of BEM.

\subsubsection{G coefficient}

Note that the case $f=F$ for (\ref{gfull}) is not singular, the second term of the integrand in
(\ref{gfull}) being zero. Therefore, we will deal only with this case. The integration of the second term
that exists when $f<F$ is similar to the first.

The divergence of $E_1(z)$ at $z\rightarrow 0$ is logarithmic and
thus integrable. Usually this means that the integration can be
performed by the Gauss method. However, since our problem is
singular due to the contact line effects, many Gauss points are
needed to attain the required accuracy in the contact line region
and a more sophisticated algorithm is necessary to get both
accuracy and speed. The analytical integration \cite{Lagier} is
used when $y=0$ i.e. when the singularity occurs. Although there
is no singularity when $y\neq 0$, the integrand varies sharply
near the point $u=-x$ when $y\ll l$. For the case $|x|<l/2$, the
integration interval can be split by the point $u=-x$ and changes
of variables can be done in both integrals to produce
\begin{eqnarray}
G_{ij}^{FF}=&{\sqrt{\rm Fo}\over 4\pi}\left\{\int\limits_{y^2/4{\rm Fo}}^{[(x+l/2)^2+y^2]/4{\rm
Fo}}{E_1(z)\over \sqrt{z-y^2/4{\rm Fo}}}\,{\rm d}z+\right.\nonumber\\
&\left.\int\limits_{y^2/4{\rm Fo}}^{[(x+l/2)^2+y^2]/4{\rm
Fo}}{E_1(z)\over \sqrt{z-y^2/4{\rm Fo}}}\,{\rm
d}z\right\}.\label{gF}
\end{eqnarray}
At first glance, no advantage is obtained because of the divergence. However, the approximation \cite{Abr}
of $E_1(z)$ for $z<1$ allows the analytical integration to be performed term by term and results in
\begin{eqnarray}
  \int_b^a{E_1(z)\over\sqrt{z-b}}\,{\rm d}z=
  -4\sqrt{b}\arctan\sqrt{a-b\over b}+2\sqrt{a-b}\,[1.422784-\log(a)+\nonumber\\
  0.333331\,(a+2b)-0.0166607\,(3a^2+4ab+8b^2)+\nonumber\\1.577134\cdot10^{-3}\,(5a^3+6a^2b+8ab^2+16b^3)- \nonumber\\
3.09843\cdot10^{-5}(35a^4+40a^3b+48a^2b^2+64ab^3+128b^4)+\nonumber\\
1.55638\cdot10^{-6}(63a^5+70a^4b+80a^3b^2+96a^2b^3+128ab^4+256b^5)],\label{E1}
\end{eqnarray}
where $b\leq a<1$ are assumed. For the remaining part of the integration interval, the 8-point Gauss
integration is performed and gives a sufficient accuracy.

The case $|x|>l/2$ should be solved similarly when the argument of
$E_1$ in (\ref{gfull}) can be less than unity somewhere in the
integration interval.

\subsubsection{H coefficient}

The value of the coefficient (\ref{gfull}) at $y=0$ is zero. While
no singularity exists when $f<F$ (this case can be integrated by
Gauss method), the function
\begin{equation}
  H_{ij}^{FF}={y\over 2\pi}\exp\left(-{y^2\over 4\,{\rm
Fo}}\right)\int\limits_{x-l/2}^{x+l/2}{\exp\left(-{u^2\over 4\,{\rm Fo}}\right)\over u^2+y^2}\;{\rm
d}u\label{hF}
\end{equation}
is discontinuous at $y=0$. The integration interval in (\ref{hF}) contains the point $u=0$ when $|x|<l/2$.
Since the integrand is an even function of $u$, this integral can be presented as
$$\int\limits_{x-l/2}^{x+l/2}\ldots=\int\limits_0^{x+l/2}\ldots+ \int\limits_0^{l/2-x}\ldots$$ and the
interval $(0,\varepsilon)$, where $0<\varepsilon\ll l/2$, can be separated out of the both integrals. The
contribution of this interval to (\ref{hF}) turns out to be
 $$H_{ij}^{FF}={1\over\pi}\exp\left(-{y^2+\varepsilon^2/4\over 4\,{\rm
Fo}}\right)\arctan{\varepsilon\over y}+\ldots$$ The discontinuity is evident now: while $y\rightarrow+0$
limit is $1/2$, the value at $y=0$ is zero and $y\rightarrow-0$ limit is $-1/2$.

After the integration over the interval $(-\varepsilon,\varepsilon)$ analytically, the Gauss method can be
employed to integrate over the remaining parts of the interval in (\ref{hF}).

\subsection{Calculation scheme}

The system (\ref{lin1}) can be simplified due to axial symmetry of the problem
$\zeta,\psi_{fj}=\zeta\psi_{f(2N_f-j)}$:
\begin{equation}
\sum\limits_{f=1}^F\sum\limits_{j=1}^{N_f}[(\zeta_{fj}/k+ \psi_{fj}v^n_{fj}/{\rm
Fo})\tilde{G}_{ij}^{Ff}-\psi_{fj}\tilde{H}_{ij}^{Ff}]=\psi_{Fi}/2,\label{fin}
\end{equation}
where $i=1\ldots N_F$, $F=1\ldots F_{max}$, $\tilde{G}_{ij}^{Ff}=G_{ij}^{Ff}+G_{i(2N_f-j)}^{Ff}$ and
$\tilde{H}_{ij}^{Ff}=H_{ij}^{Ff}+H_{i(2N_f-j)}^{Ff}$. Unfortunately, no effective time marching scheme
\cite{Breb} can be applied because of the free boundaries. The position of each node depends on time.
Therefore, it is very important that $\tilde{G}_{ij}^{Ff}$ be calculated using the coordinates of the
$i$-th point at time moment $F$ and those of $j$-th point at time moment $f$.

Our 2D BEM algorithm was tested for the fixed boundary wedge problem (for the $\Omega_L$ domain only) with
the constant heat flux boundary condition described in section \ref{sec2}. Since $\zeta$ decreases to zero
far from the contact point, integration contour can be closed at the distance $x_{max}\sim 10\sqrt{{\rm
Fo}\,t}$ from the contact point O$(0,0)$ where  $\zeta(x,y,t)$ is sufficiently small. The element lengths
grow exponentially ($d_{min}, d_{min}e^b, d_{min}e^{2b},\ldots$) from the contact point into each of the
sides of the wedge (see Fig.~\ref{Fig1}), where $b$ is fixed at 0.2. Since $x_{max}$ increases with time,
the total number of the elements also increases during the evolution of the bubble. Being an input
parameter, $d_{min}$ is adjusted slightly on each time step to provide the exponential growth law for the
elements on the interval with the fixed boundaries $(0, x_{max})$. Remeshing on each time step was
performed to comply with the free boundary nature of the main problem where the remeshing is mandatory.

The results for $\theta=\pi/8,\;\pi/4$ and $\pi/2$ are shown in Fig.~\ref{test} to be compared with the
analytical solutions \cite{IJHMT,EuLet}. It is easy to see that the method produces excellent results
except for the element closest to the contact point. The algorithm overestimates the value of $q_L$ at this
element. The error is larger for the $\pi/2$ wedge, for which $q_L\rightarrow\infty$ at the contact point.
We verified that the influence of the increase of the time and space steps on the numerical error is very
weak.

The discretization of the integration subcontours of the bubble growth problem $\partial\Omega_w$,
$\partial\Omega_d$ and $\partial\Omega_i$ follows the same exponential scheme (see Fig. \ref{bubble}) that
was used for the discretization of the wedge sides in the test example above. Since the free boundary
introduces a nonlinearity into the problem, the following iteration algorithm is needed to determine the
bubble shape on each time step \cite{Wrobel}:
\begin{enumerate}
  \item Shape of the bubble is guessed to be the same
  as on the previous time step;
  \item The variations of $v^n$ and $P_r$ along the bubble
  interface are guessed to be the same as on the previous time step;
  \item Discretization of the contours $\partial\Omega_w$,
  $\partial\Omega_d$ and $\partial\Omega_i$ is performed;
  \item Temperatures and fluxes on the contours $\partial\Omega_w$,
  $\partial\Omega_d$ and $\partial\Omega_i$ are found by solving (\ref{fin})
  for $\psi,\zeta_{L,S}$ at the time moment $F$;
  \item Volume $V$ and vapor recoil $P_r$ are calculated using
  (\ref{eqnV}) and (\ref{Pr});
  \item Bubble shape is determined for the
  calculated values of $V$ and $P_r$;
  \item If the calculated shape differs too much from that on the
  previous iteration,  the velocity of interface $v^n$ is calculated, and
   steps 3 -- 7 are repeated until the required accuracy is attained.
\end{enumerate}
As a rule, three iterations give the 0.1\% accuracy which is sufficient for our purposes.

The normal velocity of interface $v^n_{Fi}$ at the time $F$ and at node $i$ is calculated using the
expression
\begin{equation}\label{vn0}
  v^n_{Fi}=(x_{Fi}-x_{(F-1)j})n^x_{(F-1)j}+(y_{Fi}-y_{(F-1)j})n^y_{(F-1)j},
\end{equation}
where $x_{Fi}$ is the coordinate of the node $i$ at time $F$, and $j$ is the number of the node (at time
$F-1$) geometrically closest to $(x_{Fi},y_{Fi})$.

The system of Eqs. \ref{vol1} -- \ref{vol3} is solved by direct integration. The integration of the
right-hand side of (\ref{vol3}) is performed using the simple mid-point rule, because the values of $P_r$
are calculated at the mid-points (nodes) only. The subsequent integration of the right-hand sides of
Eqs.~\ref{vol1} -- \ref{vol2} is performed using the Simpson rule (to gain accuracy) for the non-equal
intervals. The trapezoidal rule turns out to be accurate enough for the calculation of volume in (\ref{V}).
For the simulation we used the parameters for water at 10 MPa pressure on the heater made of stainless
steel listed in \cite{IJHMT}.

The above described algorithm should give good results when $\int_0^1P_r(\xi){\rm d}\xi$ exists (cf.
Eq.~\ref{L}). In our case $P_r(\xi)$ can be approximated by the power function $(1-\xi)^{-2\beta}$ when
$\xi\rightarrow 1$. The exponent $\beta$, which comes from the approximation for $q_L(\xi)$, turns out to
be larger than one half (see discussion in the next section). Thus if the data were extrapolated to the
contact point $\xi=1$, this integral would diverge. We note, however, that the evaporation heat flux is
limited \cite{Carey} by a flux $q_{max}$ which is about $10^4\mbox{ MW/m}^2$ for the chosen parameters.

The $q_L$ divergence is originated from the assumption that the temperature remains constant along the
vapor-liquid interface. In reality, this assumption is violated in the very close vicinity of the contact
point where the heat flux $q_L$ is comparable to $q_{max}$. Thus we accept the following approximation for
the function $q_L(\xi)$, $\xi<1$. It is extrapolated using the power law $q_L(\xi)\propto(1-\xi)^{-\beta}$
until it reaches the value of $q_{max}$ and remains constant while $\xi$ increases to unity. This
extrapolation is used to calculate the integrals in (\ref{eqnV}) and (\ref{L}). There is no need to modify
the constant-temperature boundary condition for the heat transfer calculations because the calculated heat
flux $q_L$ remains always less than $q_{max}$.

\section{Results of the numerical calculation}

\begin{figure}[ht]
  \includegraphics[height=6cm]{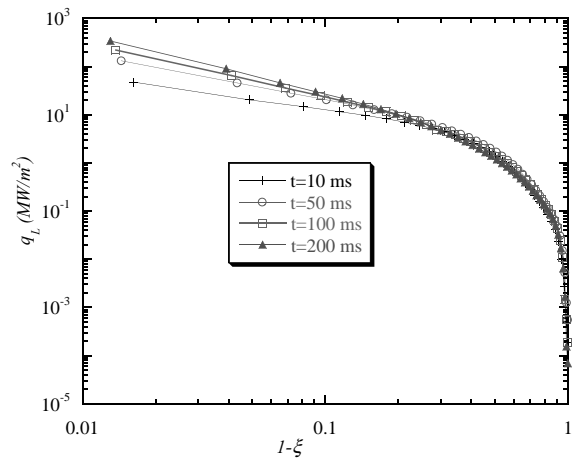}
\caption{Variation of the heat flux $q_L$ calculated for $q_0=0.05$ MW/m${}^2$ along the bubble contour for
different moments of time. The curvilinear coordinate $\xi$ varies along the bubble contour; $\xi=1$ at
the contact point. (Reprinted from \cite{IJHMT} with permission from Elsevier Science)} \label{q_L}
\end{figure}

\begin{figure}
\includegraphics[height=6cm]{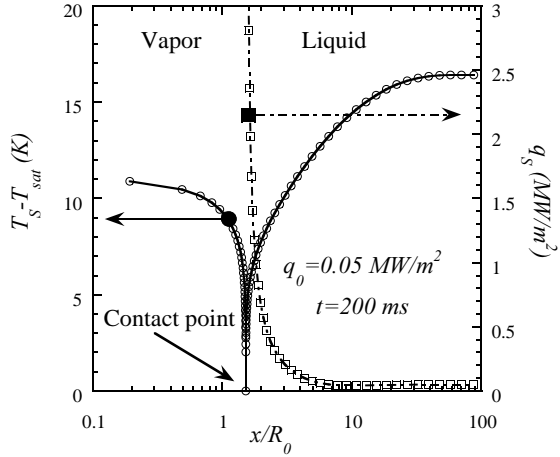}
\caption{Variation of the heat flux $q_S$ and the temperature $T_S$ along the surface of the heater for
$q_0=0.05$ MW/m${}^2$, $t=200$ ms. The point $x=0$ corresponds to the center of the bubble. (Reprinted from
\cite{IJHMT} with permission from Elsevier Science)} \label{qT_S}
\end{figure}

\begin{figure}\includegraphics[height=5cm]{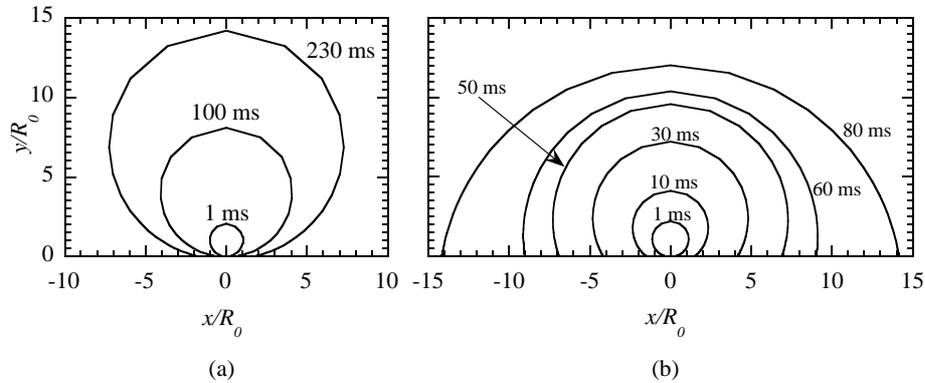}
\caption{The bubble shape shown for the different growth times.\newline a)~$q_0=0.05$~MW/m${}^2$;
b)~$q_0=0.5$~MW/m${}^2$.\newline (Reprinted from \cite{IJHMT} with permission from Elsevier
Science)}\label{shape}
\end{figure}

The calculations show that the function $q_L(\xi)$ (see Fig.~\ref{q_L})
can be described well by the above power law where $\beta\sim 1$ grows slightly with time. As expected, for
this conjugate heat conduction problem the divergence is stronger than for the wedge model,
Fig.~\ref{test}. The difference between these two cases is caused by the behavior of the heat flux $q_{S}$
in the vicinity of the contact point. While it was imposed to be uniform for the wedge, the function
$q_{S}(x)$ increases strongly near the contact point for the conjugate heat conduction, see
Fig.~\ref{qT_S}.
As expected, the value of $q_{S}$ on the liquid side in the vicinity of the contact line is very close to
$q_{L}$, the heat flux that produces evaporation on the vapor-liquid interface and that diverges on the
contact line (see Fig.~\ref{q_L}). At some distance from the bubble center $q_S$ reaches the value of
$q_0$, the flux at infinity. Note that the Fig.~\ref{qT_S} corresponds to the quasi-spherical bubble shape
(see Fig.~\ref{shape}a) and the visible bubble radius (see Fig.~\ref{bubble} for the definition) $R\sim
7R_0$, so that the heat flux $q_S$ is virtually homogeneous outside the bubble.

The variation of the temperature along the heating surface $T_S(x)$ is also shown in Fig.~\ref{qT_S}. Far
from the bubble, $T_S=T_{S}^{inf}$ has to increase with time independently of $x$ according to
(\ref{TSinf}). It decreases to $T_{sat}$ near the contact point because the temperature should be equal to
$T_{sat}$ on the whole vapor-liquid interface, according to the imposed boundary condition. Inside the dry
spot, $T_{S}$ increases with time because the heat transfer through the dry spot is blocked. It is smaller
than $T_{S}^{inf}$ while the dry spot under the bubble remains much smaller than $R$. Since $q_L$ grows
with time, at some time moment the vapor recoil pressure $P_r\propto q_L^2$ overcomes the surface tension
(cf. Eq. \ref{surf}) and the dry spot begins to grow, see Fig.~\ref{shape}b. This bubble spreading was
observed experimentally in several works, e. g. \cite{PRE}. The spreading bubble serves as a nucleus for
the formation of a continuous vapor film that separates the solid from the liquid, i. e. triggers the
boiling crisis. The heat transfer becomes blocked at a larger portion of the solid surface (i.e. dry spot)
and its temperature grows faster than $T_{S}^{inf}$. This temperature increase leads to the burnout of the
heater observed during the boiling crisis.

\section{Conclusions}

Several conclusions can be made. Our analysis of the bubble growth dynamics shows the strong coupling of
the heat conduction in the liquid and in the solid heater. Therefore, only a conjugate heat transfer
calculation of the bubble growth can properly simulate the boiling crisis. We show that BEM suits well for
such a simulation. We carried out a 2D BEM calculation. It shows that the vapor recoil can be at the
origin of the boiling crisis. We demonstrated how a vapor bubble spreads over the heating surface thus
initiating the boiling crisis.

\section*{Acknowledgements}

We gratefully acknowledge a permission of Elsevier Science B. V. to reproduce some figures from
\cite{IJHMT}. We thank our colleagues H. Lemonnier, J. Hegseth and G.-L. Lagier for the fruitful
discussions.

\end{document}